\documentclass[journal=langd5,manuscript=article]{achemso}
\DeclareSymbolFont{matha}{OML}{txmi}{m}{it}% txfonts
\DeclareMathSymbol{\varv}{\mathord}{matha}{118}
\usepackage[version=3]{mhchem} % Formula subscripts using \ce{}
\usepackage{amssymb}
\usepackage{xcolor}
\usepackage{soul,ulem}
\usepackage{physics}
\usepackage{upgreek,textgreek}
 \newcommand{\blu}[1]{{\color{black}{#1}}}
\author{Russell Kajouri}
\affiliation[ifpan]
{Institute for Computational Physics, University of Stuttgart, 70569 Stuttgart, Germany}
\author{Panagiotis E. Theodorakis}
\affiliation[ifpan]
{Institute of Physics, Polish Academy of Sciences, Al. Lotnik\'ow 32/46, 02-668 Warsaw, Poland}
 \email{panos@ifpan.edu.pl}
\author{Andrey Milchev}
\affiliation{Bulgarian Academy of Sciences, Institute of Physical Chemistry, 1113 Sofia, Bulgaria}

   \title[Durotaxis and Antidurotaxis Droplet Motion onto Gradient Gel-Substrates]
  {Durotaxis and Antidurotaxis Droplet Motion onto Gradient Gel-Substrates}

\keywords{Droplets, Wetting, Gradient Substrates, Durotaxis, Polymer Brush, Motion steering, Molecular Dynamics}

\begin{document}

%\begin{tocentry}
%\includegraphics[width=8.25cm]{FIGS/toc.pdf}
%\end{tocentry}

\begin{abstract}
The self-sustained motion of fluids on gradient substrates is 
a spectacular phenomenon, which can be employed and
controlled in applications  by carefully engineering the substrate properties. 
Here, we report on a design of a gel-substrate with stiffness
gradient, which can cause the spontaneous motion of a droplet
along (durotaxis) or to the opposite (antidurotaxis) direction of the
gradient, depending on the droplet
affinity to the substrate. By using 
extensive molecular dynamics simulations of a coarse-grained model, we
find that the mechanisms of the durotaxis and antidurotaxis droplet motion
are distinct, require the minimization of the interfacial energy between
the droplet and the substrate, and share similarities with
those mechanisms previously observed for brush
substrates with stiffness gradient. 
Moreover, durotaxis motion takes place over a wider
range of affinities and is generally more efficient (faster motion) than
antidurotaxis. 
Thus, our study points to
\blu{further} possibilities and guidelines for
realizing both antidurotaxis and durotaxis motion on the same gradient substrate
for applications in microfluidics, energy conservation, and biology.
\end{abstract}

\vspace{0.7in}

\section{INTRODUCTION}

The autonomous motion of fluids on gradient substrates has 
been observed in various contexts, for example, in the
case of microfluidics, microfabrication,
coatings, energy conversion, and biology. 
\cite{Srinivasarao2001,Chaudhury1992,Wong2011,Lagubeau2011,Prakash2008,Darhuber2005, Yao2012, Li2018, Becton2016,vandenHeuvel2007,DuChez2019,Khang2015,Ricard2023}
Moreover, both the efficiency and the direction of motion can be controlled by
carefully engineering the gradient of a substrate property. 
In the case of moving cells on 
tissues,\cite{DuChez2019,Khang2015,Lo2000,Pham2016,Lazopoulos2008} 
their motion has been attributed to gradients in
the stiffness of the underlying tissue, a phenomenon known as durotaxis. 
Inspired by biological systems, 
efforts to foster new possibilities of sustained motion
on substrates with gradually changing properties along a certain direction
have taken place, in view of the spectrum of
possible applications in diverse areas.
This also includes nano-objects of different type (\textit{e.g.} fluids,
nanosheets) on a wide range of different substrates, which
have been studied in the context of theoretical and simulation 
work,\cite{Theodorakis2017,Chang2015,Becton2014,Barnard2015,Palaia2021,Tamim2021,Bardall2020,Kajouri2023,Kajouri2023b,Huang2024} 
as well as experiments.\cite{Style2013,Hiltl2016}

The exciting aspect of durotaxis is
the autonomously sustained motion, that is no
energy supply from an external source is required for setting in
and sustaining the motion of the nano-object. 
While in connection with durotaxis, a gradient in the
stiffness is responsible for the motion, such motion
can actually be observed in other scenarios as well, for example,
when the gradient reflects changes in the pattern of the substrate.
Here, a characteristic example is rugotaxis, where a fluid motion
is caused by a gradient in the wavelength characterizing a
wavy substrate.\cite{Theodorakis2022,Hiltl2016}
Other examples include curvotaxis, that is motion attributed
to curvature changes, such as that observed in the context
of curved protein complexes at the cell.\cite{Sadhu2023}
Further possibilities, include small condensate droplets
that can move due to the presence of asymmetric pillars\cite{Feng2020},
three-dimensional (3D) capillary ratchets,\cite{Feng2021}
or pinning and depinning effects at the three-phase contact
line.\cite{Theodorakis2021} 
Interestingly, in the case of capillary ratchets, the surface tension
can play a role in determining the direction of motion,
whether this is along or against the gradient.\cite{Feng2021}
In addition, substrates with wettability gradients have been reported
as a possibility for the autonomous motion of 
liquids,\cite{Pismen2006,Wu2017,Sun2023} for example,
due to corrosion,\cite{Ricard2023} while
long-range transport has been realized by using 
electrostatic\cite{Sun2019,Jin2022} or triboelectric charges\cite{Xu2022}.
In the presence of an external energy source, motion 
is also possible, with characteristic examples being electrotaxis\cite{Jin2023}
and thermotaxis.\cite{Zhang2022} For example, in the latter
case, the motion is caused by a temperature gradient that 
requires to be maintained along the substrate by means of an external energy source.
Further examples of motion due to external sources
include motion caused by 
electrical current \cite{Dundas2009,Regan2004,Zhao2010,Kudernac2011},
charge \cite{Shklyaev2013,Fennimore2003,Bailey2008}, 
or even simple stretching\cite{Huang2014}.
Situations where droplets are chemically driven
have also been reported in the literature\cite{Santos1995,Lee2002},
as well as droplets on vibrated 
substrates\cite{Daniel2002,Brunet2007,Brunet2009,Kwon2022}
or wettability ratchets\cite{Buguin2002,Thiele2010,Noblin2009,Ni2022}.

Motivated by relevant experiments with liquid 
droplets,\cite{Style2013,Hiltl2016} we have previously
proposed and investigated by computer simulation
various substrate designs that can cause a sustained 
droplet motion.\cite{Theodorakis2017,Theodorakis2022,Kajouri2023,Kajouri2023b}
More specifically, we have proposed two designs of
brush substrates with stiffness gradient that can cause such motion 
either along or against the gradient direction.\cite{Kajouri2023,Kajouri2023b}
In the first design, the brush substrate had a constant density of
grafted polymer chains.\cite{Kajouri2023} 
In this case, the stiffness gradient was a result of changes
in the stiffness of the individual polymer chains along the gradient direction.
We have found that the droplet can move toward areas of higher stiffness (durotaxis),
where a larger number of contacts between the droplet and the substrate
can be established, due to a lower substrate roughness in these areas.
In the second design of a brush substrate, the grafted polymer chains
were fully flexible and the stiffness gradient was imposed by changing
the grafting density along a particular direction.\cite{Kajouri2023b}
In this case, the droplet could move toward softer parts of the substrate (antidurotaxis),
establishing more pair contacts as it penetrated into the substrate. 
Interestingly, the latter antidurotaxis motion might share similarities with experiments
of droplets on soft substrates with stiffness gradient, where droplet
motion was also observed from stiffer toward softer areas of the 
substrate.\cite{Style2013}
Moreover, in this case, larger droplets seem to perform
antidurotaxis motion more efficiently (faster), an effect that might not
be attributed to gravity effects due to
the weight of the droplet, as experiments were carried out
for micrometer-sized water droplets, i.e., smaller than the capillary length ($\sim$2.5~mm).

Thus far, experimental
substrates\cite{DuChez2019,Khang2015,Lo2000,Pham2016,Lazopoulos2008,Style2013}
and simulation models\cite{Theodorakis2017,Kajouri2023,Palaia2021,Chang2015,Kajouri2023b}
have mostly demonstrated either durotaxis or antidurotaxis motion for a given substrate. 
Here, building upon our previous experience with durotaxis and antidurotaxis droplet motion onto
brush substrates,\cite{Kajouri2023,Kajouri2023b} we show that a novel
gel substrate can demonstrate both antidurotaxis and durotaxis droplet motion
depending on the type of liquid. To achieve this result, a gradient in
the bonding stiffness
between the gel chemical units is used in our model to create the stiffness 
gradient along a specific direction of the gel substrate. 
Furthermore, by means of extensive molecular dynamics (MD) simulations
of a coarse-grained model, we 
elucidate the mechanisms for both the durotaxis and antidurotaxis
motions and their efficiency for a range of parameters relevant for
this substrate design. Interestingly, we observe similarities for these
mechanisms with what we have previously seen for
brush substrates.\cite{Kajouri2023,Kajouri2023b}
Thus, this may point to more universal features of such substrates
that can cause durotaxis and antidurotaxis motion of fluids,
and holds hope for the experimental realization of such substrates.
In the following, we provide details of the system, simulation model
and methodology. Then, we will present and discuss the obtained results, while  we
will draw the conclusions resulting from our investigations in the final section.

\section{MATERIALS AND METHODS}

\begin{figure}[bt!]
\centering
\includegraphics[width=\columnwidth]{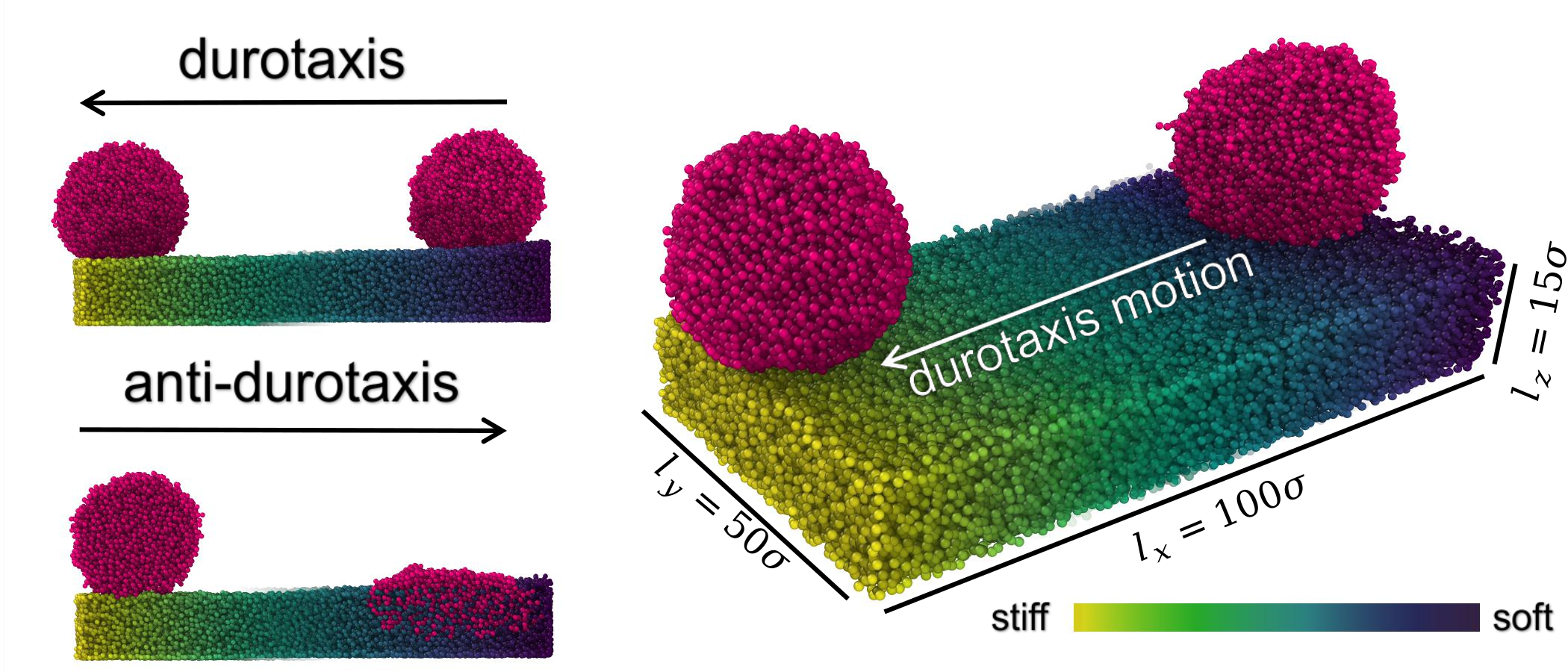}

 \caption{
Typical conformations of the substrate after equilibration with the droplet on top at an
initial and a final position with the direction of motion clearly indicated
by the arrow for both the durotaxis (upper left and right panels) 
and the antidurotaxis cases (lower left panel). 
The stiffness gradient is visually represented by the color gradient with yellow
reflecting areas of the highest stiffness and dark blue of the lowest. 
The dimensions of the gel substrate in the $x$ (gradient direction), $y$, and $z$ directions
are $l_x=100~\sigma$, $l_y=50~\sigma$, and $l_z\approx15~\sigma$, respectively.
The snapshot of the system was obtained using Ovito software.\cite{Stukowski2010}
} \label{fig:1} 
\end{figure}

The gel substrate of this study is illustrated in \textbf{Figure~\ref{fig:1}}
with typical configurations of the droplet at the beginning and the end
of successful durotaxis/antidurotaxis simulations.
In particular, the droplet remains on the top of the substrate as
it reaches the stiffest end of the substrate in the durotaxis case, while the droplet
appears to penetrate into the substrate in the case of antidurotaxis motion as
it reaches the softest end of the substrate. The length of the substrate
in the direction of the gradient is $l_x=100~\sigma$, where $\sigma$ is the length unit.
The gel substrate is supported by a
smooth and unstructured substrate and consists of beads each initially placed at the positions
of the vertices of a simple cubic lattice with unity lattice constant (expressed in units of 
$\sigma$)  with harmonic interactions between beads reaching up to
the second nearest neighbors. To realize the gradient in
the substrate stiffness, the magnitude of these interactions 
(elastic constant, $\Gamma_{\rm s}$ in units of $\varepsilon/\sigma^{2}$,
where $\varepsilon$ is the energy unit)
linearly varies with the position $x$ of the beads obtaining larger values
towards the stiffer regions of the substrate (\textbf{Figure~\ref{fig:1}}),
while the equilibrium length is
set to $1.2~\sigma$. 
The rate of change of $\Gamma_{\rm s}$ is
$0.05~\varepsilon/\sigma^{3}$ at steps of $2~\sigma$ starting from an initial 
value of $\Gamma_{\rm s} =0.5~\varepsilon/\sigma^{2}$ at the softest end of the substrate,
thus implying $\Gamma_{\rm s} =5~\varepsilon/\sigma^{2}$ at the stiffest end.
Since this particular choice was proven to be
optimal for carrying out our parametric investigation,
our results will be based on this specific substrate setup.
Once the substrate reached its equilibrium state by means of molecular
dynamics simulation (further details will be provided below), a polymer
droplet was first placed onto the softest and then the stiffest part of the substrate to examine 
the direction of motion (antidurotaxis or durotaxis). Once, the direction
of motion was identified, the decision was taken onto which end of the substrate
the droplet should be placed and an ensemble of simulations were carried out
for each set of parameters.

Nonbonded interactions between particles (beads) in the system are based on
the Lennard-Jones (LJ) potential, expressed by the relation
\begin{equation}\label{eq:LJpotential}
U_{\rm LJ}(r) = 4\varepsilon_{\rm ij} \left[  \left(\frac{\sigma_{\rm  ij}}{r}
\right)^{12} - \left(\frac{\sigma_{\rm ij}}{r}  \right)^{6}    \right].
\end{equation}
Here, $r$ is the distance between any pair of beads,
with indices ${\rm i}$ and ${\rm j}$ in \textbf{Equation~\ref{eq:LJpotential}}
reflecting the type of bead, namely ``d'' for the droplet and ``g'' for the
gel substrate. The size of the beads is the same, namely $\sigma_{\rm ij} = \sigma$.
Attractive interactions between the gel and
the droplet beads as well as among the droplet beads are used by
choosing a cutoff of $r_{\rm c}=2.5~\sigma$ for the LJ potential,
while an athermal model is used for the interactions among the gel beads.
The strength of LJ interactions between the droplet beads is set to
$\varepsilon_{\rm dd}=1.5~\varepsilon$. Different choices are considered
for the interaction strength between the polymer droplet and the gel substrate,
namely $\varepsilon_{\rm dg}=0.3-1.0~\varepsilon$, thus in practice controlling 
the affinity of the droplet to the substrate.
Finally, the droplet consists of fully flexible polymer chains
to avoid evaporation effects, which may also further complicate our analysis.
Hence, the vapor pressure is sufficiently low.\cite{Tretyakov2014}
In particular, the droplet consists of polymer chains with length 
$N_{\rm l}=10$ beads each, while the total size of the droplet is 8000 beads.
To bind the beads together in each polymer chain of the droplet
a harmonic potential was used with elastic constant $1000~\varepsilon/\sigma^2$
and equilibrium length $\sigma$.

To control the temperature of the system, $T=\varepsilon/k_B$
($k_B$ is Boltzmann's constant),
the Nos\'{e}--Hoover  thermostat was applied,\cite{Martyna1994,Cao1996}
as implemented in the HOOMD-Blue package (version 2.9.7).\cite{hoomd-blue}
The integration time step was set to $0.005~\tau$, where
$\tau=(m\sigma^{2}/\varepsilon)^{1/2}$ is the natural MD time
unit. For every set of parameters, we perform five 
simulation experiments with different initial conditions
(e.g., changing the random seed for generating the initial
velocities of the system) to statistically collect data
for the analysis of properties.
Finally, each simulation run lasts a total of $50\times10^6$
time steps, which was deemed long enough for drawing reliable
conclusions on the possibility of observing the droplet motion
and carrying out the necessary analysis of the relevant
properties.

\begin{figure}[bt!]
\centering
 \includegraphics[width=.49\textwidth]{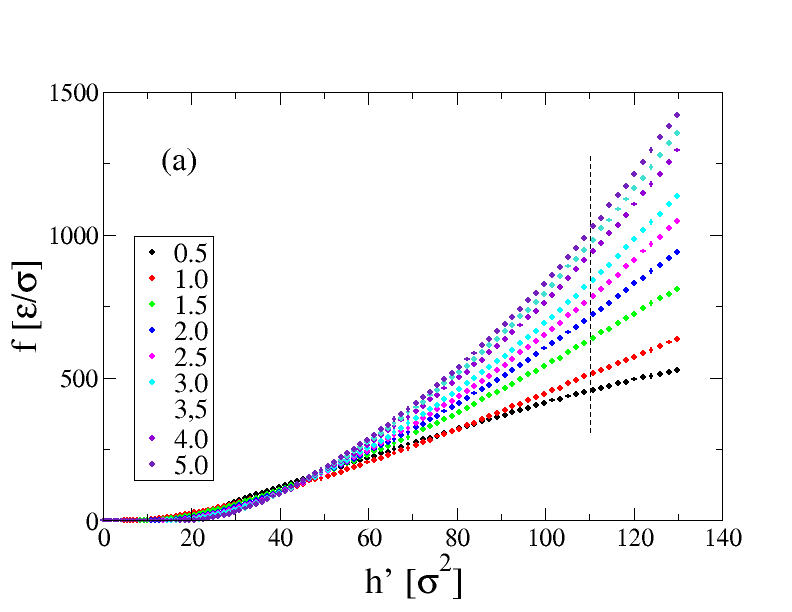}
\includegraphics[width=.49\textwidth]{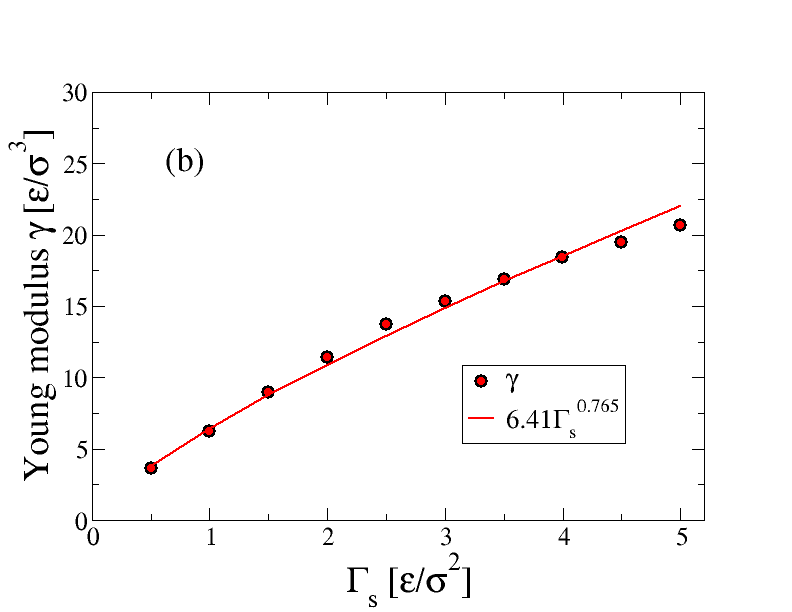}
\caption{(a) Force \textit{versus} $h'=\alpha h^{3/2}$
for different choice of the elastic constant for 
the harmonic interaction between gel beads, $\Gamma_{\rm s}$,
as indicated. According to \textbf{Equation~\ref{Eq:Hertz}}, the slope
yields the Young modulus, $\gamma$, for each case.
Here, data refer to substrates with no stiffness gradient. Hence,
$\Gamma_{\rm s}$ is constant along the substrate for each case.
The maximum value used for the fit is indicated by the
vertical dashed line.
(b) Young modulus, $\gamma$, for each substrate without gradient
\textit{versus} $\Gamma_{\rm s}$. The solid, red line 
corresponds to a power-law fit.
} \label{fig:2}
\end{figure}

Before presenting our durotaxis and antidurotaxis experiments
and their analysis, we take a step back to analyze the
stiffness of the substrate and see how this varies with
the strength of the interactions between beads used to
create the gradient. 
To perform our analysis, we consider a nanoindenter that
slowly impinges onto the gel substrate without gradient, as
has been done in a previous study in the case of protein fibrils.\cite{Poma2019} 
By recording the total
force of the substrate beads on the nanoindenter, the Young modulus, $\gamma$,
of the gel substrate can be determined similarly to
an empirical technique used to estimate the Young modulus 
in atomic-force-microscopy (AFM) nanoindentaion experiments. The Young modulus of the
nanoindenter is infinite and we therefore define each system in the limit
of the Hertzian theory.\cite{Hertz1882} 
The indenter is a sphere with a curvature radius $R_{\rm ind}$ that
slowly impinges onto the gel substrate with a velocity $u_{\rm ind}$.
Here, this velocity was the same in all nanoindentation exeriments,
i.e., data were collected every $5\times10^3$ MD time steps for a total
trajectory length of $5\times10^5$ time steps with a time step of $0.005~\tau$.
Then, the nanoindentation force, $f$, is defined by the Hertz relation
\begin{equation}\label{Eq:Hertz}
f = \alpha \gamma  h^{3/2}, 
\end{equation}
where $h$ is the penetration or nanoindentation length, $\gamma$
Young modulus, and 
\begin{equation}\label{Eq:Hertz-alpha}
\alpha = \frac{4R_{\rm ind}^{1/2}}{3(1-\nu^2)} 
\end{equation}
In our simulation experiments, the radius of the nanoindenter
was $R_{\rm ind}=5~\sigma$ and the maximum penetration depth $h_{\rm max}=10~\sigma$.
$\nu$ is the Poisson coefficient, in our case taken as 0.5,
which corresponds to a homogeneous deformation on the $x-y$ plane.
Then, the Young modulus, $\gamma$, can be determined by calculating the
slope of the curves of \textbf{Figure~\ref{fig:2}a} for 
each gel substrate without gradient but with a different value of 
the harmonic elastic constant, $\Gamma_{\rm s}$. 
By plotting the obtained Young's moduli as a function of
$\Gamma_{\rm s}$ (\textbf{Figure~\ref{fig:2}b}), we 
conclude that increasing $\Gamma_{\rm s}$ indeed results
in stiffer gel substrates. By attempting to fit a power-law
function on these data, we obtained an exponent of about 3/4 for the
relation between $\gamma$ and $\Gamma_{\rm s}$.

\section{RESULTS AND DISCUSSION}

\begin{figure}[bt!]
\centering
  \includegraphics[width=0.5\textwidth]{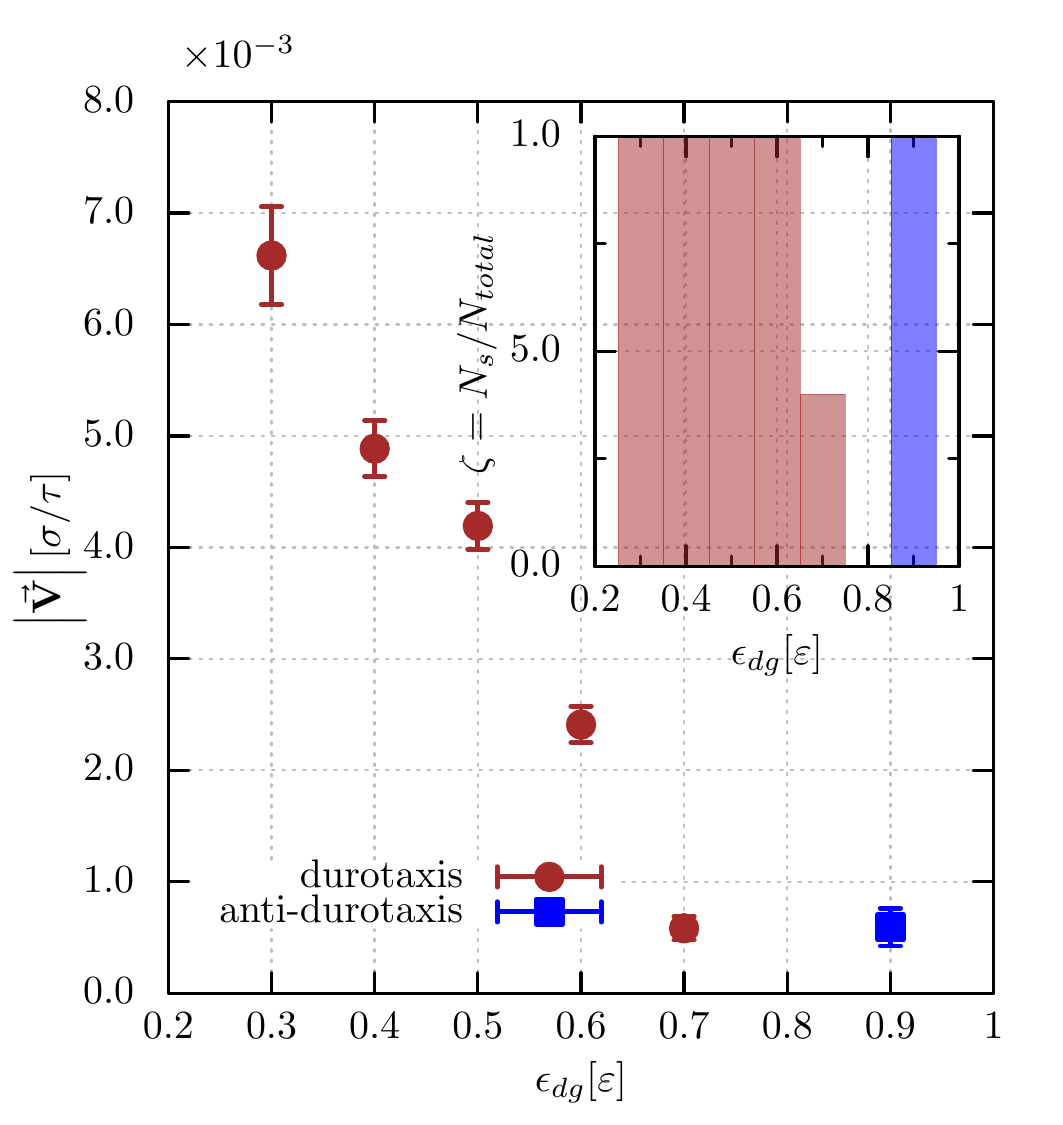}
\caption{
The average speed of the droplet as calculated from successful durotaxis/antidurotaxis
experiments ($N_{\rm s}$ is the number of the successful
cases) from a total ensemble of five ($N_{\rm total}=5$) 
trajectories for each case, as indicated.
Inset shows the probability, $\zeta=N_{\rm s}/N_{\rm total}$,
of the droplet moving from one
side of the substrate to the other. This probability for antidurotaxis
cases is illustrated by purple bars,
while that for the durotaxis cases by brown color. 
For $\varepsilon_{\rm dg}<0.8~\varepsilon$ durotaxis is observed, while
antidurotaxis was recorded for $\varepsilon_{\rm dg}=0.9~\varepsilon$.
For $\varepsilon_{\rm dg}=0.8~\varepsilon$, only partial droplet motion
was observed from each trajectory and therefore no successful cases 
are reported in the plot.
} \label{fig:3}
\end{figure}

Given the constant gradient maintained in each of our
simulation experiments, which is optimally chosen to 
facilitate our properties exploration, 
the first aspect of our research concerns the possibility of causing durotaxis
or antidurotaxis motion and the probability of such motion 
for a range of droplet--substrate affinities. 
To address this issue, a droplet is placed either on the
softest or the stiffest part of the substrate and the outcome
of the simulation is monitored. 
\textbf{Figure~\ref{fig:3}} visually summarizes our conclusions. 
For values $\varepsilon_{\rm dg}<0.2~\varepsilon$, 
the interaction between the droplet and the gel substrate
is weak. Hence, in this case the droplet
detaches from the substrate due to the thermal fluctuations
and this case deserves no further consideration here.
Durotaxis motion takes place when
$0.2~\varepsilon<\varepsilon_{\rm dg}<0.8~\varepsilon$.
For this range of affinity strength between the
droplet and the substrate, we observe that the droplet moves
from softer to stiffer parts of the gel substrate
covering its full length in the $x$ direction,
a manifestation of successful durotaxis motion for the droplet. 
While for $0.3~\varepsilon\leq\varepsilon_{\rm dg}\leq0.6~\varepsilon$
the probability that the droplet successfully moves from the softest
to the stiffest side of the substrate is 1.0 as calculated from an
ensemble of five different trajectories for each affinity
case,
this probability becomes less than unity when $\varepsilon_{\rm dg}=0.7~\varepsilon$.
Moreover, we were able to only detect partial motion along
the substrate, when $\varepsilon_{\rm dg}=0.8~\varepsilon$,
reporting threrefore this case as unsuccessful. 
This provides a first indication that the droplet motion may become
less effective for larger values of $\varepsilon_{\rm dg}$. 
Indeed, this is
corroborated by monitoring the average velocity
of the droplet for different values $\varepsilon_{\rm dg}$ (\textbf{Figure~\ref{fig:3}}),
which clearly indicates that an increased affinity between the
droplet and the substrate will lead to a smaller
average durotaxis velocity.
Further increase of the affinity, namely $\varepsilon_{\rm dg}=0.9~\varepsilon$,
leads to successful antidurotaxis motion. In this case, the droplet
reached the softest
part of the gel substrate and the recorded average velocity was 
of the
same magnitude as in the durotaxis case with $\varepsilon_{\rm dg}=0.7~\varepsilon$. Finally, antidurotaxis motion for
$\varepsilon_{\rm dg}=\varepsilon$ was observed, but in this
case the droplet was not able to cover the full distance
from the one to the other side of the gel substrate
for any of our five trajectories and therefore this case
was considered unsuccessful, as was the case of partial durotaxis 
droplet motion for $\varepsilon_{\rm dg}=0.8~\varepsilon$.
The above observations may allow us to conclude that both
durotaxis and antidurotaxis motions are possible on the
same substrate. Since this takes place by varying the 
droplet--substrate affinity in our simulation,
we may argue that the direction of motion
eventually depends on the choice of liquid for the droplet.  
Also, durotaxis motion on gel substrates
is overall more efficient than the antidurotaxis motion, 
especially when the droplet--substrate affinity is lower.

\begin{figure}[bt!]
\centering
 \includegraphics[width=.49\textwidth]{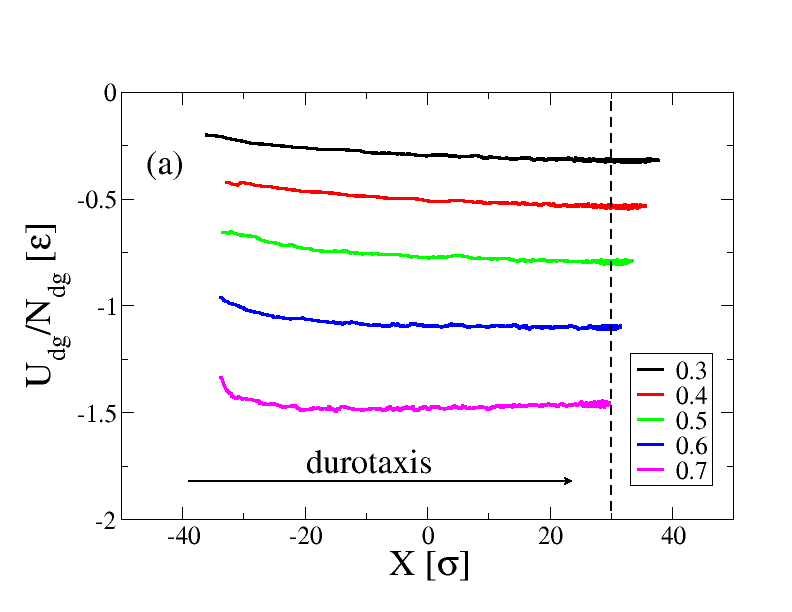}
%\hspace=0.5cm
\includegraphics[width=.49\textwidth]{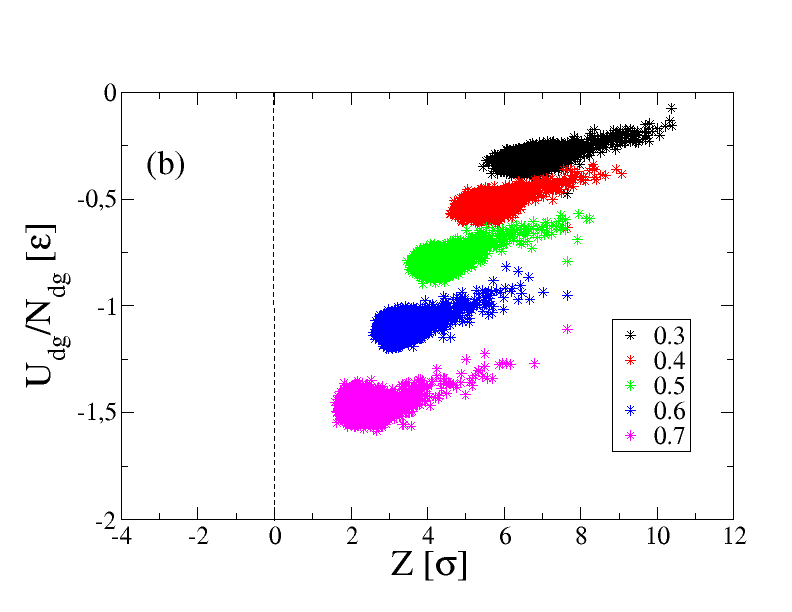} 
\caption{(a) Interfacial energy of the droplet normalized
by the number of substrate--droplet bead pairs
as a function of the $X$ 
coordinate of the center-of-mass of the droplet in the $x$
direction for a range of different durotaxis cases with
different $\varepsilon_{\rm dg}$, as indicated. 
The vertical dashed line indicates the $X$ position
for the center-of-mass considered for determining the
successful translocation of the droplet toward the
stiffest end of the substrate.
(b) The normalized
interfacial energy is plotted against the coordinate
of the center-of-mass of the droplet in the $z$ direction.
The vertical dashed line denotes the position of 
gel's surface, calculated by the inflection point of
the density profile of the gel, as is done in our
previous study.\cite{Kajouri2023b}
} \label{fig:4}
\end{figure}

\begin{figure}[bt!]
\centering
 \includegraphics[width=.49\textwidth]{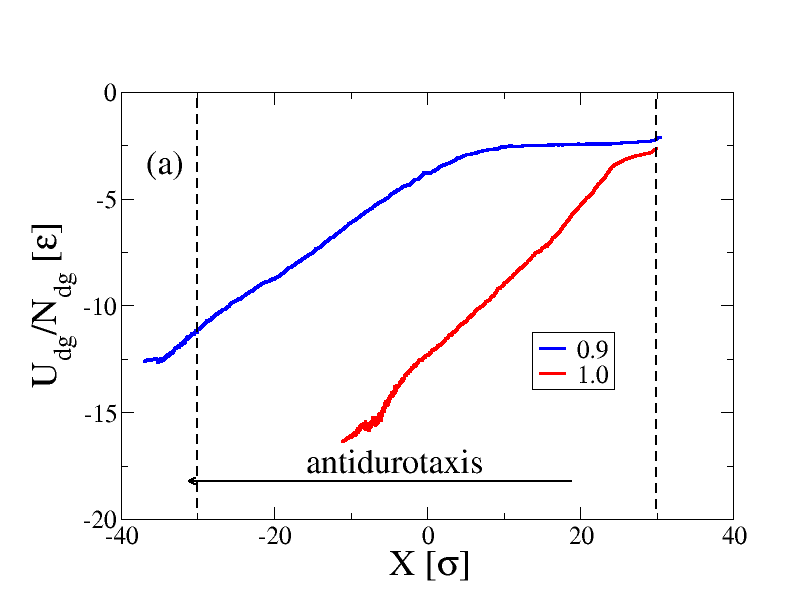}
\includegraphics[width=.49\textwidth]{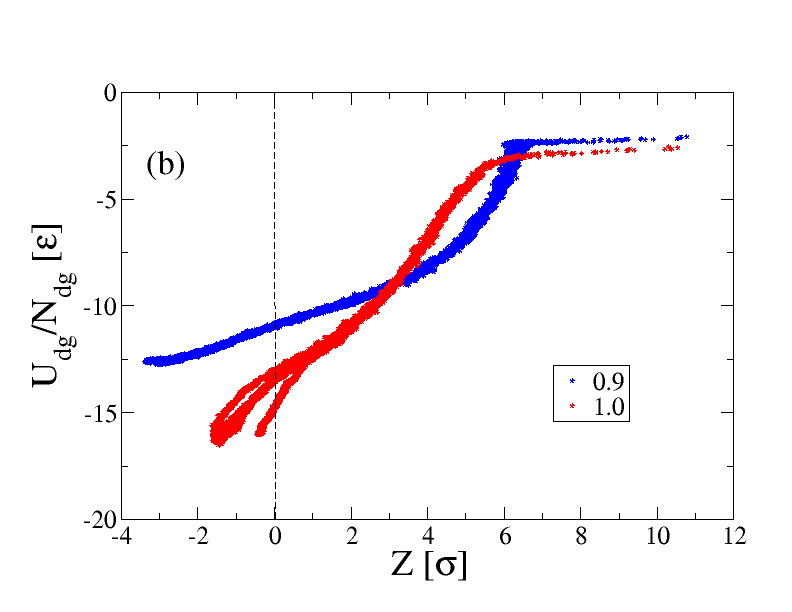}
\caption{(a) Interfacial energy of the droplet normalized
by the number of substrate--droplet bead pairs
as a function of the $X$ 
coordinate of the center-of-mass of the droplet in the $x$
direction for a range of different antidurotaxis cases with
different $\varepsilon_{\rm dg}$, as indicated. 
(b) The interfacial energy is plotted against the coordinate
of the center-of-mass of the droplet in the $z$ direction.
} \label{fig:5}
\end{figure}

As in our previous studies,\cite{Theodorakis2017,Theodorakis2022,Kajouri2023,Kajouri2023b} 
we attempted to identify the
driving force for both antidurotaxis and durotaxis cases.
$X$ in \textbf{Figure~\ref{fig:4}} 
indicates the coordinate of the center-of-mass of the 
droplet in the $x$ direction with the zero value corresponding to the
center of the gel substrate. $Z$ is the coordinate of the 
center-of-mass of the droplet in the $z$ direction with the
zero indicating the position of the substrate boundary, 
which was determined through the inflection point in
the density profile of each substrate as
done in our previous work.~\cite{Kajouri2023b}
Moreover, the peculiarities of the gel--droplet interface have been explored recently in detail.\cite{Vishnu2024}
On the basis of our analysis for the durotaxis cases,
we observe that the interfacial energy
between the droplet and the substrate decreases as a function
of the center-of-mass position of the droplet in 
both the $x$ (\textbf{Figure~\ref{fig:4}a}) and the $z$ directions (\textbf{Figure~\ref{fig:4}b}),
which suggests that the droplet establish a larger number of
contacts with the gel as the it moves along the substrate
(see also \textbf{Movie 1} in the \textbf{Supporting Information}).
As a result, the droplet is more strongly attracted by the gel as
it moves toward the stiffer parts, which results in a
decrease in the position $Z$ of the center-of-mass of the droplet, but
with the droplet however remaining on top of the substrate.
Moreover, we observe that the slope in the energy reduction of
the interfacial energy as a function of the position $X$ of the
center-of-mass of the droplet is larger for smaller values
of the attraction strength
$\varepsilon_{\rm dg}$ (\textbf{Figure~\ref{fig:4}a}), which reflects
the conclusions relating to the average velocity of the droplet presented
in \textbf{Figure~\ref{fig:3}}, that is a lower adhesion of the droplet
to the gel substrate offers a more efficient (in terms of droplet speed)
durotaxis motion.
This motion mechanism of the droplet shares similarities with
the durotaxis motion previously observed on brush substrates,\cite{Kajouri2023}
where the droplet moves to the areas of smaller surface fluctuations
of the substrate, that is substrate parts of lower roughness.

The results of \textbf{Figure~\ref{fig:4}} for the durotaxis cases
can be compared with those for the antidurotaxis cases
presented in \textbf{Figure~\ref{fig:5}}. 
Notably, we observe that the interfacial energy is
much more reduced for the antidurotaxis cases in
comparison with the durotaxis ones.
More importantly, we also see that the droplet 
penetrates deeper into the substrate 
in the case of antidurotaxis droplet motion
and the center-of-mass of the droplet eventually lies
below the top of the substrate as the antidurotaxis motion completes
(see also \textbf{Movie 2} of the \textbf{Supporting Information}).
This mechanism is therefore more similar to the 
one observed in the case of antidurotaxis motion for
brush substrates with
gradient in the grafting density of the polymer
chains.\cite{Kajouri2023b} In this case, the minimization
of the interfacial energy was due to the penetration
of the droplet onto the brush substrate.
For this reason, the droplet motion is much less efficient
than that in the case of durotaxis simulations, since the droplet
faces a larger resistance in carrying out the motion along
the substrate by bypassing the gel beads.

\begin{figure}[bt!]
\centering \includegraphics[width=.49\textwidth]{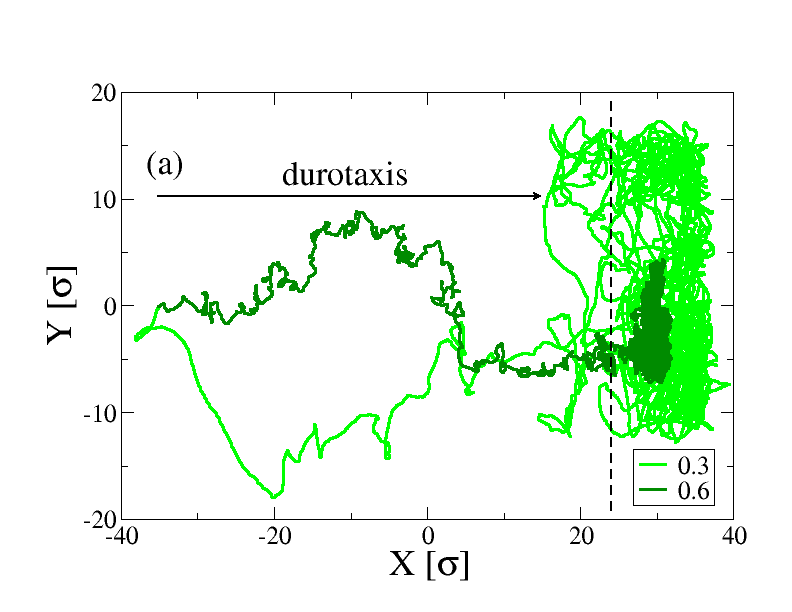}
\includegraphics[width=.49\textwidth]{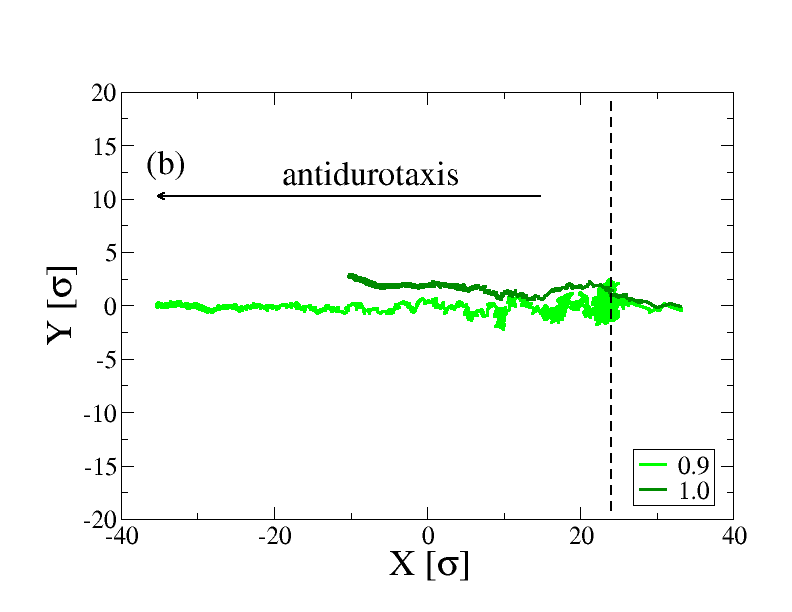} 
\caption{Typical trajectories for (a) durotaxis  and (b)
antidurotaxis cases for values of $\varepsilon_{\rm dg}$, as indicated. 
The points of the trajectory
are collected by tracking the center-of-mass of the droplet
on the $x-y$ plane.
} \label{fig:6}
\end{figure}

Finally, we monitored the trajectories of the center-of-mass
of the droplet onto the $x-y$ plane (\textbf{Figure~\ref{fig:6}}). 
A different behavior of the droplet motion is observed
between durotaxis and antidurotaxis cases.
In particular, we see that the droplet motion is
more influenced by thermal fluctuations 
as indicated by the lateral motion in the $y$ direction
in the case of durotaxis (\textbf{Figure~\ref{fig:6}a}). The droplet
clearly initially moves at a higher instantaneous speed 
towards the stiffer areas and then slightly
slows down. This pattern of motion is observed
for both the lowest and the highest affinity between the droplet and
the substrate, which may indicate that the affinity might play a lesser
role in determining the exact trajectory of the particle. 
The weakening effect of the gradient effect on the droplet
velocity as the droplet reaches the ever stiffer parts
of the substrate has been thus far observed in all
previous durotaxis/antidurotaxis studies.\cite{Theodorakis2017,Kajouri2023,Kajouri2023b}
In the case of antidurotaxis experiments (\textbf{Figure~\ref{fig:6}b}),
the droplet appears to only move in the $x$ direction
with minimal lateral (diffusive) motion in the $y$ direction,
which may suggest that the motion in this case is dominated by the 
droplet--substrate interactions. This takes place to a larger degree
as the droplet moves to the softer parts of the substrate.
Hence, we can see that the droplet motion fundamentally differs in the case
of antidurotaxis and durotaxis cases, with the antidurotaxis motion providing
a more certain path for the trajectory of the droplet moving
along the substrate during the
simulation experiments.

\section{CONCLUSIONS}
In this study, we have proposed and investigated a 
novel substrate design based on a gel material. 
Importantly, we have been able to demonstrate that
durotaxis and antidurotaxis motion of a droplet
is possible on the same substrate and the 
direction of motion only depends on the 
fluid. 
To our knowledge, this is the first time that this 
possibility is realized for gel substrates.
As in the case of durotaxis onto brush substrates\cite{Kajouri2023,Kajouri2023b},
we have found that the minimization of the interfacial
energy between the droplet and the substrate
is the dominant driving force responsible for
the motion of the droplet. This takes
place by the substantial penetration of the
substrate by the droplet in the case of antidurotaxis
or the droplet motion towards areas with smaller
surface fluctuations on the top of the gel
in the case of durotaxis. 
As a result, the trajectories of the droplet
motion appear to be more diffusive in the
durotaxis cases than in the antidurotaxis
cases, where in the latter
the droplet motion is hindered by the 
gel units. 
Moreover, recent experiments \cite{Zhao2024_arxiv} 
have reported on the 
spontaneous droplet motion on soft, gel substrates with
stiffness gradient created by varying the degree of 
cross-linking
in the gel. In this case, results have pointed out to
the minimization of the interfacial energy between
the substrate and the droplet as the driving
force for the durotaxis motion of the droplet, as in the case of simulation experiments here and in previous studies.\cite{Theodorakis2017,Kajouri2023,Kajouri2023b}
We have also found that durotaxis takes place
for a wide range of droplet--substrate affinities with
lower affinities leading to more efficient durotaxis motion,
while fully successful antidurotaxis motion 
has only been observed for a high value of
droplet--substrate affinity. 

Our study provides further evidence that both 
durotaxis and antidurotaxis motion can be 
realized on the same gel substrate. Thus,
we anticipate that our work highlights the new
venues of possibilities in the autonomous motion
of fluids based on gradient gel-substrates
and provides insights into the motion of 
droplets driven by stiffness gradients, thus 
enhancing our understanding of similar phenomena, encountered
in nature.

\begin{acknowledgement}
Authors thank Jan \v{Z}idek for helpful discussions.
This research has been supported by the National Science Centre, Poland, under grant
No. 2019/35/B/ST3/03426. A. M. acknowledges support by COST (European Cooperation in
Science and Technology [See http://www.cost.eu and https://www.fni.bg] and its 
Bulgarian partner FNI/MON under KOST-11). 
We gratefully acknowledge Polish high-performance
computing infrastructure PLGrid
(HPC Centers: ACK Cyfronet AGH)
for providing computer facilities and 
support within computational
grant no. PLG/2023/016607.
\end{acknowledgement}

\begin{suppinfo}

Movie1.mp4: Movie illustrates the droplet durotaxis motion ($\varepsilon_{\rm dg}=0.3~\varepsilon$).
\\
\noindent
Movie2.mp4: Movie illustrates the droplet antidurotaxis motion ($\varepsilon_{\rm dg}=0.9~\varepsilon$).
\end{suppinfo}

\bibliography{bib_gel}

\providecommand{\latin}[1]{#1}
\makeatletter
\providecommand{\doi}
  {\begingroup\let\do\@makeother\dospecials
  \catcode`\{=1 \catcode`\}=2 \doi@aux}
\providecommand{\doi@aux}[1]{\endgroup\texttt{#1}}
\makeatother
\providecommand*\mcitethebibliography{\thebibliography}
\csname @ifundefined\endcsname{endmcitethebibliography}
  {\let\endmcitethebibliography\endthebibliography}{}
\begin{mcitethebibliography}{69}
\providecommand*\natexlab[1]{#1}
\providecommand*\mciteSetBstSublistMode[1]{}
\providecommand*\mciteSetBstMaxWidthForm[2]{}
\providecommand*\mciteBstWouldAddEndPuncttrue
  {\def\EndOfBibitem{\unskip.}}
\providecommand*\mciteBstWouldAddEndPunctfalse
  {\let\EndOfBibitem\relax}
\providecommand*\mciteSetBstMidEndSepPunct[3]{}
\providecommand*\mciteSetBstSublistLabelBeginEnd[3]{}
\providecommand*\EndOfBibitem{}
\mciteSetBstSublistMode{f}
\mciteSetBstMaxWidthForm{subitem}{(\alph{mcitesubitemcount})}
\mciteSetBstSublistLabelBeginEnd
  {\mcitemaxwidthsubitemform\space}
  {\relax}
  {\relax}

\bibitem[Srinivasarao \latin{et~al.}(2001)Srinivasarao, Collings, Philips, and
  Patel]{Srinivasarao2001}
Srinivasarao,~M.; Collings,~D.; Philips,~A.; Patel,~S. Three-dimensionally
  ordered array of air bubbles in a polymer film. \emph{Science} \textbf{2001},
  \emph{292}, 79--83\relax
\mciteBstWouldAddEndPuncttrue
\mciteSetBstMidEndSepPunct{\mcitedefaultmidpunct}
{\mcitedefaultendpunct}{\mcitedefaultseppunct}\relax
\EndOfBibitem
\bibitem[Chaudhury and Whitesides(1992)Chaudhury, and
  Whitesides]{Chaudhury1992}
Chaudhury,~M.~K.; Whitesides,~G.~M. How to Make Water Run Uphill.
  \emph{Science} \textbf{1992}, \emph{256}, 1539--1541\relax
\mciteBstWouldAddEndPuncttrue
\mciteSetBstMidEndSepPunct{\mcitedefaultmidpunct}
{\mcitedefaultendpunct}{\mcitedefaultseppunct}\relax
\EndOfBibitem
\bibitem[Wong \latin{et~al.}(2011)Wong, Kang, Tang, Smythe, Hatton, Grinthal,
  and Aizenberg]{Wong2011}
Wong,~T.-S.; Kang,~S.~H.; Tang,~S. K.~Y.; Smythe,~E.~J.; Hatton,~B.~D.;
  Grinthal,~A.; Aizenberg,~J. Bioinspired self-repairing slippery surfaces with
  pressure-stable omniphobicity. \emph{Nature} \textbf{2011}, \emph{477},
  443--447\relax
\mciteBstWouldAddEndPuncttrue
\mciteSetBstMidEndSepPunct{\mcitedefaultmidpunct}
{\mcitedefaultendpunct}{\mcitedefaultseppunct}\relax
\EndOfBibitem
\bibitem[Lagubeau \latin{et~al.}(2011)Lagubeau, Le~Merrer, Clanet, and
  Qu\'er\'e]{Lagubeau2011}
Lagubeau,~G.; Le~Merrer,~M.; Clanet,~C.; Qu\'er\'e,~D. Leidenfrost on a
  ratchet. \emph{Nat. Phys.} \textbf{2011}, \emph{7}, 395--398\relax
\mciteBstWouldAddEndPuncttrue
\mciteSetBstMidEndSepPunct{\mcitedefaultmidpunct}
{\mcitedefaultendpunct}{\mcitedefaultseppunct}\relax
\EndOfBibitem
\bibitem[Prakash \latin{et~al.}(2008)Prakash, Qu\'er\'e, and Bush]{Prakash2008}
Prakash,~M.; Qu\'er\'e,~D.; Bush,~J.~W. Surface tension transport of prey by
  feeding shorebirds: The capillary ratchet. \emph{Science} \textbf{2008},
  \emph{320}, 931--934\relax
\mciteBstWouldAddEndPuncttrue
\mciteSetBstMidEndSepPunct{\mcitedefaultmidpunct}
{\mcitedefaultendpunct}{\mcitedefaultseppunct}\relax
\EndOfBibitem
\bibitem[Darhuber and Troian(2005)Darhuber, and Troian]{Darhuber2005}
Darhuber,~A.; Troian,~S. Principles of microfluidic actuation by modulation of
  surface stresses. \emph{Annu. Rev. Fluid Mech.} \textbf{2005}, \emph{37},
  425--455\relax
\mciteBstWouldAddEndPuncttrue
\mciteSetBstMidEndSepPunct{\mcitedefaultmidpunct}
{\mcitedefaultendpunct}{\mcitedefaultseppunct}\relax
\EndOfBibitem
\bibitem[Yao and Bowick(2012)Yao, and Bowick]{Yao2012}
Yao,~Z.; Bowick,~M.~J. {Self-propulsion of droplets by spatially-varying
  surface topography}. \emph{Soft Matter} \textbf{2012}, \emph{8},
  1142--1145\relax
\mciteBstWouldAddEndPuncttrue
\mciteSetBstMidEndSepPunct{\mcitedefaultmidpunct}
{\mcitedefaultendpunct}{\mcitedefaultseppunct}\relax
\EndOfBibitem
\bibitem[Li \latin{et~al.}(2018)Li, Yan, Fichthorn, and Yu]{Li2018}
Li,~H.; Yan,~T.; Fichthorn,~K.~A.; Yu,~S. {Dynamic contact angles and
  mechanisms of motion of water droplets moving on nano-pillared
  superhydrophobic surfaces: A molecular dynamics simulation study}.
  \emph{Langmuir} \textbf{2018}, \emph{34}, 9917--9926\relax
\mciteBstWouldAddEndPuncttrue
\mciteSetBstMidEndSepPunct{\mcitedefaultmidpunct}
{\mcitedefaultendpunct}{\mcitedefaultseppunct}\relax
\EndOfBibitem
\bibitem[Becton and Wang(2016)Becton, and Wang]{Becton2016}
Becton,~M.; Wang,~X. {Controlling nanoflake motion using stiffness gradients on
  hexagonal boron nitride}. \emph{RSC Adv.} \textbf{2016}, \emph{6},
  51205--51210\relax
\mciteBstWouldAddEndPuncttrue
\mciteSetBstMidEndSepPunct{\mcitedefaultmidpunct}
{\mcitedefaultendpunct}{\mcitedefaultseppunct}\relax
\EndOfBibitem
\bibitem[van~den Heuvel and Dekker(2007)van~den Heuvel, and
  Dekker]{vandenHeuvel2007}
van~den Heuvel,~M. G.~L.; Dekker,~C. Motor proteins at work for nanotechnology.
  \emph{Science} \textbf{2007}, \emph{317}, 333--336\relax
\mciteBstWouldAddEndPuncttrue
\mciteSetBstMidEndSepPunct{\mcitedefaultmidpunct}
{\mcitedefaultendpunct}{\mcitedefaultseppunct}\relax
\EndOfBibitem
\bibitem[DuChez \latin{et~al.}(2019)DuChez, Doyle, Dimitriadis, and
  Yamada]{DuChez2019}
DuChez,~B.~J.; Doyle,~A.~D.; Dimitriadis,~E.~K.; Yamada,~K.~M. Durotaxis by
  human cancer cells. \emph{Biophys. J.} \textbf{2019}, \emph{116},
  670--683\relax
\mciteBstWouldAddEndPuncttrue
\mciteSetBstMidEndSepPunct{\mcitedefaultmidpunct}
{\mcitedefaultendpunct}{\mcitedefaultseppunct}\relax
\EndOfBibitem
\bibitem[Khang(2015)]{Khang2015}
Khang,~G. Evolution of gradient concept for the application of regenerative
  medicine. \emph{Biosurface Biotribology} \textbf{2015}, \emph{1},
  202--213\relax
\mciteBstWouldAddEndPuncttrue
\mciteSetBstMidEndSepPunct{\mcitedefaultmidpunct}
{\mcitedefaultendpunct}{\mcitedefaultseppunct}\relax
\EndOfBibitem
\bibitem[Ricard \latin{et~al.}(2023)Ricard, Restagno, Jang, Lansac, and
  Raspaud]{Ricard2023}
Ricard,~A.; Restagno,~F.; Jang,~Y.~H.; Lansac,~Y.; Raspaud,~E. Corrosion-driven
  droplet wetting on iron nanolayers. \emph{Sci. Rep.} \textbf{2023},
  \emph{13}, 18288\relax
\mciteBstWouldAddEndPuncttrue
\mciteSetBstMidEndSepPunct{\mcitedefaultmidpunct}
{\mcitedefaultendpunct}{\mcitedefaultseppunct}\relax
\EndOfBibitem
\bibitem[Lo \latin{et~al.}(2000)Lo, Wang, Dembo, and Wang]{Lo2000}
Lo,~C.-M.; Wang,~H.-B.; Dembo,~M.; Wang,~Y.-L. Cell movement is guided by the
  rigidity of the substrate. \emph{Biophys. J.} \textbf{2000}, \emph{79},
  144--152\relax
\mciteBstWouldAddEndPuncttrue
\mciteSetBstMidEndSepPunct{\mcitedefaultmidpunct}
{\mcitedefaultendpunct}{\mcitedefaultseppunct}\relax
\EndOfBibitem
\bibitem[Pham \latin{et~al.}(2016)Pham, Xue, {Del Campo}, and
  Salierno]{Pham2016}
Pham,~J.~T.; Xue,~L.; {Del Campo},~A.; Salierno,~M. {Guiding cell migration
  with microscale stiffness patterns and undulated surfaces}. \emph{Acta
  Biomaterialia} \textbf{2016}, \emph{38}, 106--115\relax
\mciteBstWouldAddEndPuncttrue
\mciteSetBstMidEndSepPunct{\mcitedefaultmidpunct}
{\mcitedefaultendpunct}{\mcitedefaultseppunct}\relax
\EndOfBibitem
\bibitem[Lazopoulos and Stamenovi{\'{c}}(2008)Lazopoulos, and
  Stamenovi{\'{c}}]{Lazopoulos2008}
Lazopoulos,~K.~A.; Stamenovi{\'{c}},~D. {Durotaxis as an elastic stability
  phenomenon}. \emph{J. Biomech.} \textbf{2008}, \emph{41}, 1289--1294\relax
\mciteBstWouldAddEndPuncttrue
\mciteSetBstMidEndSepPunct{\mcitedefaultmidpunct}
{\mcitedefaultendpunct}{\mcitedefaultseppunct}\relax
\EndOfBibitem
\bibitem[Theodorakis \latin{et~al.}(2017)Theodorakis, Egorov, and
  Milchev]{Theodorakis2017}
Theodorakis,~P.~E.; Egorov,~S.~A.; Milchev,~A. Stiffness-guided motion of a
  droplet on a solid substrate. \emph{J. Chem. Phys.} \textbf{2017},
  \emph{146}, 244705\relax
\mciteBstWouldAddEndPuncttrue
\mciteSetBstMidEndSepPunct{\mcitedefaultmidpunct}
{\mcitedefaultendpunct}{\mcitedefaultseppunct}\relax
\EndOfBibitem
\bibitem[Chang \latin{et~al.}(2015)Chang, Zhang, Guo, Guo, and Gao]{Chang2015}
Chang,~T.; Zhang,~H.; Guo,~Z.; Guo,~X.; Gao,~H. {Nanoscale directional motion
  towards regions of stiffness}. \emph{Phys. Rev. Lett.} \textbf{2015},
  \emph{114}, 015504\relax
\mciteBstWouldAddEndPuncttrue
\mciteSetBstMidEndSepPunct{\mcitedefaultmidpunct}
{\mcitedefaultendpunct}{\mcitedefaultseppunct}\relax
\EndOfBibitem
\bibitem[Becton and Wang(2014)Becton, and Wang]{Becton2014}
Becton,~M.; Wang,~X. {Thermal gradients on graphene to drive nanoflake motion}.
  \emph{J. Chem. Theory Comput.} \textbf{2014}, \emph{10}, 722--730\relax
\mciteBstWouldAddEndPuncttrue
\mciteSetBstMidEndSepPunct{\mcitedefaultmidpunct}
{\mcitedefaultendpunct}{\mcitedefaultseppunct}\relax
\EndOfBibitem
\bibitem[Barnard(2015)]{Barnard2015}
Barnard,~A.~S. {Nanoscale locomotion without fuel}. \emph{Nature}
  \textbf{2015}, \emph{519}, 37--38\relax
\mciteBstWouldAddEndPuncttrue
\mciteSetBstMidEndSepPunct{\mcitedefaultmidpunct}
{\mcitedefaultendpunct}{\mcitedefaultseppunct}\relax
\EndOfBibitem
\bibitem[Palaia \latin{et~al.}(2021)Palaia, Paraschiv, Debets, Storm, and
  \v{S}ari\'c]{Palaia2021}
Palaia,~I.; Paraschiv,~A.; Debets,~V.~E.; Storm,~C.; \v{S}ari\'c,~A. Durotaxis
  of Passive Nanoparticles on Elastic Membranes. \emph{ACS Nano} \textbf{2021},
  \emph{15}, 15794--15802\relax
\mciteBstWouldAddEndPuncttrue
\mciteSetBstMidEndSepPunct{\mcitedefaultmidpunct}
{\mcitedefaultendpunct}{\mcitedefaultseppunct}\relax
\EndOfBibitem
\bibitem[Tamim and Bostwick(2021)Tamim, and Bostwick]{Tamim2021}
Tamim,~S.~I.; Bostwick,~J.~B. Model of spontaneous droplet transport on a soft
  viscoelastic substrate with nonuniform thickness. \emph{Phys. Rev. E}
  \textbf{2021}, \emph{104}, 034611\relax
\mciteBstWouldAddEndPuncttrue
\mciteSetBstMidEndSepPunct{\mcitedefaultmidpunct}
{\mcitedefaultendpunct}{\mcitedefaultseppunct}\relax
\EndOfBibitem
\bibitem[Bardall \latin{et~al.}(2020)Bardall, Chen, Daniels, and
  Shearer]{Bardall2020}
Bardall,~A.; Chen,~S.-Y.; Daniels,~K.~E.; Shearer,~M. {Gradient-induced droplet
  motion over soft solids}. \emph{IMA J. Appl. Math} \textbf{2020}, \emph{85},
  495--512\relax
\mciteBstWouldAddEndPuncttrue
\mciteSetBstMidEndSepPunct{\mcitedefaultmidpunct}
{\mcitedefaultendpunct}{\mcitedefaultseppunct}\relax
\EndOfBibitem
\bibitem[Kajouri \latin{et~al.}(2023)Kajouri, Theodorakis, Deuar, Bennacer,
  Židek, Egorov, and Milchev]{Kajouri2023}
Kajouri,~R.; Theodorakis,~P.~E.; Deuar,~P.; Bennacer,~R.; Židek,~J.;
  Egorov,~S.~A.; Milchev,~A. Unidirectional Droplet Propulsion onto Gradient
  Brushes without External Energy Supply. \emph{Langmuir} \textbf{2023},
  \emph{39}, 2818--2828\relax
\mciteBstWouldAddEndPuncttrue
\mciteSetBstMidEndSepPunct{\mcitedefaultmidpunct}
{\mcitedefaultendpunct}{\mcitedefaultseppunct}\relax
\EndOfBibitem
\bibitem[Kajouri \latin{et~al.}(2023)Kajouri, Theodorakis, Židek, and
  Milchev]{Kajouri2023b}
Kajouri,~R.; Theodorakis,~P.~E.; Židek,~J.; Milchev,~A. Antidurotaxis Droplet
  Motion onto Gradient Brush Substrates. \emph{Langmuir} \textbf{2023},
  \emph{39}, 15285--15296\relax
\mciteBstWouldAddEndPuncttrue
\mciteSetBstMidEndSepPunct{\mcitedefaultmidpunct}
{\mcitedefaultendpunct}{\mcitedefaultseppunct}\relax
\EndOfBibitem
\bibitem[Huang \latin{et~al.}(2024)Huang, Gu, Li, Xiang, Liao, Jiang, Ji, and
  Shen]{Huang2024}
Huang,~Z.; Gu,~C.; Li,~J.; Xiang,~P.; Liao,~Y.; Jiang,~B.-P.; Ji,~S.;
  Shen,~X.-C. Surface-Initiated Polymerization with an Initiator Gradient: A
  Monte Carlo Simulation. \emph{Polymers} \textbf{2024}, \emph{16}, 1203\relax
\mciteBstWouldAddEndPuncttrue
\mciteSetBstMidEndSepPunct{\mcitedefaultmidpunct}
{\mcitedefaultendpunct}{\mcitedefaultseppunct}\relax
\EndOfBibitem
\bibitem[Style \latin{et~al.}(2013)Style, Che, Park, Weon, Je, Hyland, German,
  Power, Wilen, Wettlaufer, and Dufresne]{Style2013}
Style,~R.~W.; Che,~Y.; Park,~S.~J.; Weon,~B.~M.; Je,~J.~H.; Hyland,~C.;
  German,~G.~K.; Power,~M.~P.; Wilen,~L.~A.; Wettlaufer,~J.~S.; Dufresne,~E.~R.
  Patterning droplets with durotaxis. \emph{Proc. Natl. Acad. Sci. U.S.A.}
  \textbf{2013}, \emph{110}, 12541--12544\relax
\mciteBstWouldAddEndPuncttrue
\mciteSetBstMidEndSepPunct{\mcitedefaultmidpunct}
{\mcitedefaultendpunct}{\mcitedefaultseppunct}\relax
\EndOfBibitem
\bibitem[Hiltl and B\"oker(2016)Hiltl, and B\"oker]{Hiltl2016}
Hiltl,~S.; B\"oker,~A. {Wetting Phenomena on (Gradient) Wrinkle Substrates}.
  \emph{Langmuir} \textbf{2016}, \emph{32}, 8882--8888\relax
\mciteBstWouldAddEndPuncttrue
\mciteSetBstMidEndSepPunct{\mcitedefaultmidpunct}
{\mcitedefaultendpunct}{\mcitedefaultseppunct}\relax
\EndOfBibitem
\bibitem[Theodorakis \latin{et~al.}(2002)Theodorakis, Egorov, and
  Milchev]{Theodorakis2022}
Theodorakis,~P.~E.; Egorov,~S.~A.; Milchev,~A. Rugotaxis: Droplet motion
  without external energy supply. \emph{EPL} \textbf{2002}, \emph{137},
  43002\relax
\mciteBstWouldAddEndPuncttrue
\mciteSetBstMidEndSepPunct{\mcitedefaultmidpunct}
{\mcitedefaultendpunct}{\mcitedefaultseppunct}\relax
\EndOfBibitem
\bibitem[Sadhu \latin{et~al.}(2023)Sadhu, Luciano, Xi, Martinez-Torres,
  Schr{\"o}der, Blum, Tarantola, Peni{\v c}, Igli{\v c}, Beta, Steinbock,
  Bodenschatz, Ladoux, Gabriele, and Gov]{Sadhu2023}
Sadhu,~R.~K.; Luciano,~M.; Xi,~W.; Martinez-Torres,~C.; Schr{\"o}der,~M.;
  Blum,~C.; Tarantola,~M.; Peni{\v c},~S.; Igli{\v c},~A.; Beta,~C.;
  Steinbock,~O.; Bodenschatz,~E.; Ladoux,~B.; Gabriele,~S.; Gov,~N.~S. A
  minimal physical model for curvotaxis driven by curved protein complexes at
  the cell's leading edge. \emph{bioRxiv} \textbf{2023}, doi:
  10.1101/2023.04.19.537490\relax
\mciteBstWouldAddEndPuncttrue
\mciteSetBstMidEndSepPunct{\mcitedefaultmidpunct}
{\mcitedefaultendpunct}{\mcitedefaultseppunct}\relax
\EndOfBibitem
\bibitem[Feng \latin{et~al.}(2020)Feng, Delannoy, Malod, Zheng, Qu\'er\'e, and
  Wang]{Feng2020}
Feng,~S.; Delannoy,~J.; Malod,~A.; Zheng,~H.; Qu\'er\'e,~D.; Wang,~Z.
  Tip-induced flipping of droplets on Janus pillars: From local reconfiguration
  to global transport. \emph{Sci. Adv.} \textbf{2020}, \emph{6}, eabb5440\relax
\mciteBstWouldAddEndPuncttrue
\mciteSetBstMidEndSepPunct{\mcitedefaultmidpunct}
{\mcitedefaultendpunct}{\mcitedefaultseppunct}\relax
\EndOfBibitem
\bibitem[Feng \latin{et~al.}(2021)Feng, Zhu, Zheng, Zhan, Chen, Li, Wang, Yao,
  Liu, and Wang]{Feng2021}
Feng,~S.; Zhu,~P.; Zheng,~H.; Zhan,~H.; Chen,~C.; Li,~J.; Wang,~L.; Yao,~X.;
  Liu,~Y.; Wang,~Z. Three dimensional capillary ratchet-induced liquid
  directional steering. \emph{Science} \textbf{2021}, \emph{373},
  1344--1348\relax
\mciteBstWouldAddEndPuncttrue
\mciteSetBstMidEndSepPunct{\mcitedefaultmidpunct}
{\mcitedefaultendpunct}{\mcitedefaultseppunct}\relax
\EndOfBibitem
\bibitem[Theodorakis \latin{et~al.}(2021)Theodorakis, Amirfazli, Hu, and
  Che]{Theodorakis2021}
Theodorakis,~P.~E.; Amirfazli,~A.; Hu,~B.; Che,~Z. Droplet Control Based on
  Pinning and Substrate Wettability. \emph{Langmuir} \textbf{2021}, \emph{37},
  4248--4255\relax
\mciteBstWouldAddEndPuncttrue
\mciteSetBstMidEndSepPunct{\mcitedefaultmidpunct}
{\mcitedefaultendpunct}{\mcitedefaultseppunct}\relax
\EndOfBibitem
\bibitem[Pismen and Thiele(2006)Pismen, and Thiele]{Pismen2006}
Pismen,~L.~M.; Thiele,~U. {Asymptotic theory for a moving droplet driven by a
  wettability gradient}. \emph{Phys. Fluids} \textbf{2006}, \emph{18},
  042104\relax
\mciteBstWouldAddEndPuncttrue
\mciteSetBstMidEndSepPunct{\mcitedefaultmidpunct}
{\mcitedefaultendpunct}{\mcitedefaultseppunct}\relax
\EndOfBibitem
\bibitem[Wu \latin{et~al.}(2017)Wu, Zhu, Cao, Zhang, Wu, Liang, Chai, and
  Liu]{Wu2017}
Wu,~H.; Zhu,~K.; Cao,~B.; Zhang,~Z.; Wu,~B.; Liang,~L.; Chai,~G.; Liu,~A. Smart
  design of wettability-patterned gradients on substrate-independent coated
  surfaces to control unidirectional spreading of droplets. \emph{Soft Matter}
  \textbf{2017}, \emph{13}, 2995--3002\relax
\mciteBstWouldAddEndPuncttrue
\mciteSetBstMidEndSepPunct{\mcitedefaultmidpunct}
{\mcitedefaultendpunct}{\mcitedefaultseppunct}\relax
\EndOfBibitem
\bibitem[Sun and Jing(2023)Sun, and Jing]{Sun2023}
Sun,~L.; Jing,~D. {Directional self-motion of nanodroplets driven by controlled
  surface wetting gradients}. \emph{Phys. Fluids} \textbf{2023}, \emph{35},
  052009\relax
\mciteBstWouldAddEndPuncttrue
\mciteSetBstMidEndSepPunct{\mcitedefaultmidpunct}
{\mcitedefaultendpunct}{\mcitedefaultseppunct}\relax
\EndOfBibitem
\bibitem[Sun \latin{et~al.}(2019)Sun, Wang, Li, Zhang, Ye, Cui, Chen, Wang,
  Butt, Vollmer, and Deng]{Sun2019}
Sun,~Q.; Wang,~D.; Li,~Y.; Zhang,~J.; Ye,~S.; Cui,~J.; Chen,~L.; Wang,~Z.;
  Butt,~H.~J.; Vollmer,~D.; Deng,~X. Surface charge printing for programmed
  droplet transport. \emph{Nat. Mater.} \textbf{2019}, \emph{18},
  936--941\relax
\mciteBstWouldAddEndPuncttrue
\mciteSetBstMidEndSepPunct{\mcitedefaultmidpunct}
{\mcitedefaultendpunct}{\mcitedefaultseppunct}\relax
\EndOfBibitem
\bibitem[Jin \latin{et~al.}(2022)Jin, Xu, Zhang, Li, Sun, Yang, Liu, Mao, and
  Wang]{Jin2022}
Jin,~Y.; Xu,~W.; Zhang,~H.; Li,~R.; Sun,~J.; Yang,~S.; Liu,~M.; Mao,~H.;
  Wang,~Z. Electrostatic tweezer for droplet manipulation. \emph{Proc. Natl.
  Acad. Sci. U.S.A.} \textbf{2022}, \emph{119}, e2105459119\relax
\mciteBstWouldAddEndPuncttrue
\mciteSetBstMidEndSepPunct{\mcitedefaultmidpunct}
{\mcitedefaultendpunct}{\mcitedefaultseppunct}\relax
\EndOfBibitem
\bibitem[Xu \latin{et~al.}(2022)Xu, Jin, Li, Song, Gao, Zhang, Wang, Cui, Yan,
  and Wang]{Xu2022}
Xu,~W.; Jin,~Y.; Li,~W.; Song,~Y.; Gao,~S.; Zhang,~B.; Wang,~L.; Cui,~M.;
  Yan,~X.; Wang,~Z. Triboelectric wetting for continuous droplet transport.
  \emph{Sci. Adv.} \textbf{2022}, \emph{8}, eade2085\relax
\mciteBstWouldAddEndPuncttrue
\mciteSetBstMidEndSepPunct{\mcitedefaultmidpunct}
{\mcitedefaultendpunct}{\mcitedefaultseppunct}\relax
\EndOfBibitem
\bibitem[Jin \latin{et~al.}(0)Jin, Liu, Xu, Sun, Huang, Yang, Yang, Wang, Lam,
  Li, and Wang]{Jin2023}
Jin,~Y.; Liu,~X.; Xu,~W.; Sun,~P.; Huang,~S.; Yang,~S.; Yang,~X.; Wang,~Q.;
  Lam,~R. H.~W.; Li,~R.; Wang,~Z. Charge-Powered Electrotaxis for Versatile
  Droplet Manipulation. \emph{ACS Nano} \textbf{0}, \emph{0}, null\relax
\mciteBstWouldAddEndPuncttrue
\mciteSetBstMidEndSepPunct{\mcitedefaultmidpunct}
{\mcitedefaultendpunct}{\mcitedefaultseppunct}\relax
\EndOfBibitem
\bibitem[Zhang \latin{et~al.}(2022)Zhang, Li, Fang, Lin, Zhao, Li, Liu, Chen,
  Lv, and Feng]{Zhang2022}
Zhang,~K.; Li,~J.; Fang,~W.; Lin,~C.; Zhao,~J.; Li,~Z.; Liu,~Y.; Chen,~S.;
  Lv,~C.; Feng,~X.-Q. An energy-conservative many-body dissipative particle
  dynamics model for thermocapillary drop motion. \emph{Phys. Fluids}
  \textbf{2022}, \emph{34}, 052011\relax
\mciteBstWouldAddEndPuncttrue
\mciteSetBstMidEndSepPunct{\mcitedefaultmidpunct}
{\mcitedefaultendpunct}{\mcitedefaultseppunct}\relax
\EndOfBibitem
\bibitem[Dundas \latin{et~al.}(2009)Dundas, McEniry, and Todorov]{Dundas2009}
Dundas,~D.; McEniry,~E.~J.; Todorov,~T.~N. Current-driven atomic waterwheels.
  \emph{Nat. Nanotechnol.} \textbf{2009}, \emph{4}, 99--102\relax
\mciteBstWouldAddEndPuncttrue
\mciteSetBstMidEndSepPunct{\mcitedefaultmidpunct}
{\mcitedefaultendpunct}{\mcitedefaultseppunct}\relax
\EndOfBibitem
\bibitem[Regan \latin{et~al.}(2004)Regan, Aloni, Ritchie, Dahmen, and
  Zettl]{Regan2004}
Regan,~B.~C.; Aloni,~S.; Ritchie,~R.~O.; Dahmen,~U.; Zettl,~A. Carbon nanotubes
  as nanoscale mass conveyors. \emph{Nature} \textbf{2004}, \emph{428},
  924\relax
\mciteBstWouldAddEndPuncttrue
\mciteSetBstMidEndSepPunct{\mcitedefaultmidpunct}
{\mcitedefaultendpunct}{\mcitedefaultseppunct}\relax
\EndOfBibitem
\bibitem[Zhao \latin{et~al.}(2010)Zhao, Huang, Wei, and Zhu]{Zhao2010}
Zhao,~J.; Huang,~J.-Q.; Wei,~F.; Zhu,~J. Mass transportation mechanism in
  electric-biased carbon nanotubes. \emph{Nano Lett.} \textbf{2010}, \emph{10},
  4309--4315\relax
\mciteBstWouldAddEndPuncttrue
\mciteSetBstMidEndSepPunct{\mcitedefaultmidpunct}
{\mcitedefaultendpunct}{\mcitedefaultseppunct}\relax
\EndOfBibitem
\bibitem[Kudernac \latin{et~al.}(2011)Kudernac, Ruangsupapichat, Parschau,
  Maci\'a, Katsonis, Harutyunyan, Ernst, and Feringa]{Kudernac2011}
Kudernac,~T.; Ruangsupapichat,~N.; Parschau,~M.; Maci\'a,~B.; Katsonis,~N.;
  Harutyunyan,~S.~R.; Ernst,~K.-H.; Feringa,~B.~L. Electrically driven
  directional motion of a four-wheeled molecule on a metal surface.
  \emph{Nature} \textbf{2011}, \emph{479}, 208--211\relax
\mciteBstWouldAddEndPuncttrue
\mciteSetBstMidEndSepPunct{\mcitedefaultmidpunct}
{\mcitedefaultendpunct}{\mcitedefaultseppunct}\relax
\EndOfBibitem
\bibitem[Shklyaev \latin{et~al.}(2013)Shklyaev, Mockensturm, and
  Crespi]{Shklyaev2013}
Shklyaev,~O.~E.; Mockensturm,~E.; Crespi,~V.~H. Theory of carbomorph cycles.
  \emph{Phys. Rev. Lett.} \textbf{2013}, \emph{110}, 156803\relax
\mciteBstWouldAddEndPuncttrue
\mciteSetBstMidEndSepPunct{\mcitedefaultmidpunct}
{\mcitedefaultendpunct}{\mcitedefaultseppunct}\relax
\EndOfBibitem
\bibitem[Fennimore \latin{et~al.}(2003)Fennimore, Yuzvinsky, Han, Fuhrer,
  Cumings, and Zettl]{Fennimore2003}
Fennimore,~A.~M.; Yuzvinsky,~T.~D.; Han,~W.-Q.; Fuhrer,~M.~S.; Cumings,~J.;
  Zettl,~A. Rotational actuators based on carbon nanotubes. \emph{Nature}
  \textbf{2003}, \emph{424}, 408--410\relax
\mciteBstWouldAddEndPuncttrue
\mciteSetBstMidEndSepPunct{\mcitedefaultmidpunct}
{\mcitedefaultendpunct}{\mcitedefaultseppunct}\relax
\EndOfBibitem
\bibitem[Bailey \latin{et~al.}(2008)Bailey, Amanatidis, and
  Lambert]{Bailey2008}
Bailey,~S. W.~D.; Amanatidis,~I.; Lambert,~C.~J. Carbon nanotube electron
  windmills: A novel design for nanomotors. \emph{Phys. Rev. Lett.}
  \textbf{2008}, \emph{100}, 256802\relax
\mciteBstWouldAddEndPuncttrue
\mciteSetBstMidEndSepPunct{\mcitedefaultmidpunct}
{\mcitedefaultendpunct}{\mcitedefaultseppunct}\relax
\EndOfBibitem
\bibitem[Huang \latin{et~al.}(2014)Huang, Zhu, and Li]{Huang2014}
Huang,~Y.; Zhu,~S.; Li,~T. {Directional transport of molecular mass on graphene
  by straining}. \emph{Extreme Mech. Lett.} \textbf{2014}, \emph{1},
  83--89\relax
\mciteBstWouldAddEndPuncttrue
\mciteSetBstMidEndSepPunct{\mcitedefaultmidpunct}
{\mcitedefaultendpunct}{\mcitedefaultseppunct}\relax
\EndOfBibitem
\bibitem[Santos; and Ondarquhus(1995)Santos;, and Ondarquhus]{Santos1995}
Santos;,~F.~D.; Ondarquhus,~T. {Free-Running Droplets}. \emph{Phys. Rev. Lett.}
  \textbf{1995}, \emph{75}, 2972\relax
\mciteBstWouldAddEndPuncttrue
\mciteSetBstMidEndSepPunct{\mcitedefaultmidpunct}
{\mcitedefaultendpunct}{\mcitedefaultseppunct}\relax
\EndOfBibitem
\bibitem[Lee \latin{et~al.}(2002)Lee, Kwok, and Laibinis]{Lee2002}
Lee,~S.~W.; Kwok,~D.~Y.; Laibinis,~P.~E. {Chemical influences on
  adsorption-mediated self-propelled drop movement}. \emph{Phys. Rev. E}
  \textbf{2002}, \emph{65}, 9\relax
\mciteBstWouldAddEndPuncttrue
\mciteSetBstMidEndSepPunct{\mcitedefaultmidpunct}
{\mcitedefaultendpunct}{\mcitedefaultseppunct}\relax
\EndOfBibitem
\bibitem[Daniel and Chaudhury(2002)Daniel, and Chaudhury]{Daniel2002}
Daniel,~S.; Chaudhury,~M.~K. Rectified motion of liquid drops on gradient
  surfaces induced by vibration. \textbf{2002}, \emph{18}, 3404--3407\relax
\mciteBstWouldAddEndPuncttrue
\mciteSetBstMidEndSepPunct{\mcitedefaultmidpunct}
{\mcitedefaultendpunct}{\mcitedefaultseppunct}\relax
\EndOfBibitem
\bibitem[Brunet \latin{et~al.}(2007)Brunet, Eggers, and Deegan]{Brunet2007}
Brunet,~P.; Eggers,~J.; Deegan,~R.~D. {Vibration-induced climbing of drops}.
  \emph{Phys. Rev. Lett.} \textbf{2007}, \emph{99}, 3--6\relax
\mciteBstWouldAddEndPuncttrue
\mciteSetBstMidEndSepPunct{\mcitedefaultmidpunct}
{\mcitedefaultendpunct}{\mcitedefaultseppunct}\relax
\EndOfBibitem
\bibitem[Brunet \latin{et~al.}(2009)Brunet, Eggers, and Deegan]{Brunet2009}
Brunet,~P.; Eggers,~J.; Deegan,~R.~D. {Motion of a drop driven by substrate
  vibrations}. \emph{Eur. Phys. J.: Spec. Top.} \textbf{2009}, \emph{166},
  11--14\relax
\mciteBstWouldAddEndPuncttrue
\mciteSetBstMidEndSepPunct{\mcitedefaultmidpunct}
{\mcitedefaultendpunct}{\mcitedefaultseppunct}\relax
\EndOfBibitem
\bibitem[Kwon \latin{et~al.}(2023)Kwon, Kim, Kim, Kim, and Kang]{Kwon2022}
Kwon,~O.~K.; Kim,~J.~M.; Kim,~H.~W.; Kim,~K.~S.; Kang,~J.~W. A Study on
  Nanosensor Based on Graphene Nanoflake Transport on Graphene Nanoribbon Using
  Edge Vibration. \emph{J. Electr. Eng. Technol.} \textbf{2023}, \emph{18},
  663--668\relax
\mciteBstWouldAddEndPuncttrue
\mciteSetBstMidEndSepPunct{\mcitedefaultmidpunct}
{\mcitedefaultendpunct}{\mcitedefaultseppunct}\relax
\EndOfBibitem
\bibitem[Buguin \latin{et~al.}(2002)Buguin, Talini, and Silberzan]{Buguin2002}
Buguin,~A.; Talini,~L.; Silberzan,~P. {Ratchet-like topological structures for
  the control of microdrops}. \emph{Appl. Phys. A: Mater. Sci. Process.}
  \textbf{2002}, \emph{75}, 207--212\relax
\mciteBstWouldAddEndPuncttrue
\mciteSetBstMidEndSepPunct{\mcitedefaultmidpunct}
{\mcitedefaultendpunct}{\mcitedefaultseppunct}\relax
\EndOfBibitem
\bibitem[Thiele and John(2010)Thiele, and John]{Thiele2010}
Thiele,~U.; John,~K. {Transport of free surface liquid films and drops by
  external ratchets and self-ratcheting mechanisms}. \emph{Chem. Phys.}
  \textbf{2010}, \emph{375}, 578--586\relax
\mciteBstWouldAddEndPuncttrue
\mciteSetBstMidEndSepPunct{\mcitedefaultmidpunct}
{\mcitedefaultendpunct}{\mcitedefaultseppunct}\relax
\EndOfBibitem
\bibitem[Noblin \latin{et~al.}(2009)Noblin, Kofman, and Celestini]{Noblin2009}
Noblin,~X.; Kofman,~R.; Celestini,~F. {Ratchetlike motion of a shaken drop}.
  \emph{Phys. Rev. Lett.} \textbf{2009}, \emph{102}, 1--4\relax
\mciteBstWouldAddEndPuncttrue
\mciteSetBstMidEndSepPunct{\mcitedefaultmidpunct}
{\mcitedefaultendpunct}{\mcitedefaultseppunct}\relax
\EndOfBibitem
\bibitem[Ni \latin{et~al.}(2022)Ni, Song, Li, Lu, Jiang, and Li]{Ni2022}
Ni,~E.; Song,~L.; Li,~Z.; Lu,~G.; Jiang,~Y.; Li,~H. Unidirectional
  self-actuation transport of a liquid metal nanodroplet in a two-plate
  confinement microchannel. \emph{Nanoscale Adv.} \textbf{2022}, \emph{4},
  2752--2761\relax
\mciteBstWouldAddEndPuncttrue
\mciteSetBstMidEndSepPunct{\mcitedefaultmidpunct}
{\mcitedefaultendpunct}{\mcitedefaultseppunct}\relax
\EndOfBibitem
\bibitem[Stukowski(2010)]{Stukowski2010}
Stukowski,~A. Visualization and analysis of atomistic simulation data with
  OVITO–the Open Visualization Tool. \emph{Modelling Simul. Mater. Sci. Eng.}
  \textbf{2010}, \emph{18}, 015012\relax
\mciteBstWouldAddEndPuncttrue
\mciteSetBstMidEndSepPunct{\mcitedefaultmidpunct}
{\mcitedefaultendpunct}{\mcitedefaultseppunct}\relax
\EndOfBibitem
\bibitem[Tretyakov and M{\"{u}}ller(2014)Tretyakov, and
  M{\"{u}}ller]{Tretyakov2014}
Tretyakov,~N.; M{\"{u}}ller,~M. {Directed transport of polymer drops on
  vibrating superhydrophobic substrates: a molecular dynamics study.}
  \emph{Soft matter} \textbf{2014}, \emph{10}, 4373--86\relax
\mciteBstWouldAddEndPuncttrue
\mciteSetBstMidEndSepPunct{\mcitedefaultmidpunct}
{\mcitedefaultendpunct}{\mcitedefaultseppunct}\relax
\EndOfBibitem
\bibitem[Martyna \latin{et~al.}(1994)Martyna, Tobias, and Klein]{Martyna1994}
Martyna,~G.~J.; Tobias,~D.~J.; Klein,~M.~L. {Constant pressure molecular
  dynamics algorithms}. \emph{J. Chem. Phys.} \textbf{1994}, \emph{101},
  4177--4189\relax
\mciteBstWouldAddEndPuncttrue
\mciteSetBstMidEndSepPunct{\mcitedefaultmidpunct}
{\mcitedefaultendpunct}{\mcitedefaultseppunct}\relax
\EndOfBibitem
\bibitem[Cao and Martyna(1996)Cao, and Martyna]{Cao1996}
Cao,~J.; Martyna,~G.~J. {Adiabatic path integral molecular dynamics methods.
  II. Algorithms}. \emph{J. Chem. Phys.} \textbf{1996}, \emph{104},
  2028--2035\relax
\mciteBstWouldAddEndPuncttrue
\mciteSetBstMidEndSepPunct{\mcitedefaultmidpunct}
{\mcitedefaultendpunct}{\mcitedefaultseppunct}\relax
\EndOfBibitem
\bibitem[Anderson \latin{et~al.}(2020)Anderson, Glaser, and
  Glotzer]{hoomd-blue}
Anderson,~J.~A.; Glaser,~J.; Glotzer,~S.~C. HOOMD-blue: A Python package for
  high-performance molecular dynamics and hard particle Monte Carlo
  simulations. \emph{Comput. Mater. Sci.} \textbf{2020}, \emph{173},
  109363\relax
\mciteBstWouldAddEndPuncttrue
\mciteSetBstMidEndSepPunct{\mcitedefaultmidpunct}
{\mcitedefaultendpunct}{\mcitedefaultseppunct}\relax
\EndOfBibitem
\bibitem[Poma \latin{et~al.}(2019)Poma, Guzman, Li, and Theodorakis]{Poma2019}
Poma,~A.~B.; Guzman,~H.~V.; Li,~M.~S.; Theodorakis,~P.~E. Mechanical and
  thermodynamic properties of \textalpha{}\textbeta{}42,
  \textalpha{}\textbeta{}40, and \textalpha{}-synuclein fibrils: a
  coarse-grained method to complement experimental studies. \emph{Beilstein J.
  Nanotechnol.} \textbf{2019}, \emph{10}, 500--513\relax
\mciteBstWouldAddEndPuncttrue
\mciteSetBstMidEndSepPunct{\mcitedefaultmidpunct}
{\mcitedefaultendpunct}{\mcitedefaultseppunct}\relax
\EndOfBibitem
\bibitem[Hertz(1882)]{Hertz1882}
Hertz,~H. Ueber die Berührung fester elastischer Körper. \emph{Journal für
  die reine und angewandte Mathematik} \textbf{1882}, \emph{1882},
  156--171\relax
\mciteBstWouldAddEndPuncttrue
\mciteSetBstMidEndSepPunct{\mcitedefaultmidpunct}
{\mcitedefaultendpunct}{\mcitedefaultseppunct}\relax
\EndOfBibitem
\bibitem[Vishnu \latin{et~al.}(2024)Vishnu, Linder, Seiffert, and
  Schmid]{Vishnu2024}
Vishnu,~J.~A.; Linder,~T.~G.; Seiffert,~S.; Schmid,~F. Structure and Dynamic
  Evolution of Interfaces between Polymer Solutions and Gels and Polymer
  Interdiffusion: A Molecular Dynamics Study. \emph{Macromolecules}
  \textbf{2024}, \emph{57}, 5545--5559\relax
\mciteBstWouldAddEndPuncttrue
\mciteSetBstMidEndSepPunct{\mcitedefaultmidpunct}
{\mcitedefaultendpunct}{\mcitedefaultseppunct}\relax
\EndOfBibitem
\bibitem[Zhao \latin{et~al.}(2024)Zhao, Qian, and Xu]{Zhao2024_arxiv}
Zhao,~W.; Qian,~W.; Xu,~Q. Spontaneous motion of liquid droplets on soft
  gradient surfaces. \emph{ArXiv} \textbf{2024}, \relax
\mciteBstWouldAddEndPunctfalse
\mciteSetBstMidEndSepPunct{\mcitedefaultmidpunct}
{}{\mcitedefaultseppunct}\relax
\EndOfBibitem
\end{mcitethebibliography}

\end{document}